\documentclass[11pt,a4paper]{article}
\usepackage{ae,aecompl}
\usepackage[T1]{fontenc}
\usepackage[latin9]{inputenc}
\usepackage{color}
\usepackage{textcomp}
\usepackage{amsmath}
\usepackage{amssymb}
\usepackage{graphicx}
\usepackage{esint}

\makeatletter

\pdfoutput=0 % if your are submitting a pdflatex (i.e. if you have
\usepackage{jheppub}% for details on the use of the package, please
\usepackage{etoolbox}% http://ctan.org/pkg/etoolbox
    
    \patchcmd{\maketitle}{\@fpheader}{}{}{}

\usepackage{aecompl}

\usepackage{amsfonts}

\usepackage{bbm}

\title{\boldmath Boundary conditions for General Relativity in three-dimensional spacetimes, integrable systems and the KdV/mKdV hierarchies}

\author[a,b]{Emilio Ojeda}
\author[a]{and Alfredo P\'{e}rez}

\affiliation[a]{Centro de Estudios Cient\'{i}ficos (CECs), Avenida Arturo Prat 514, Valdivia,
	Chile.}
\affiliation[b]{Departamento de F\'isica, Universidad de Concepci\'on, Casilla 160-C, Concepci\'on, Chile.}

\emailAdd{ojeda@cecs.cl}
\emailAdd{aperez@cecs.cl}

\preprint{CECS-PHY-19/02}

\abstract{We present a new set of boundary conditions for General Relativity
	on AdS$_3$, where the dynamics of the boundary degrees of freedom are described
	by two independent left and right members of the Gardner hierarchy of integrable equations, also known as the ``mixed KdV-mKdV'' hierarchy. This integrable system
	has the very special property that simultaneously combines both, the Korteweg-de
	Vries (KdV) and modified Korteweg-de Vries (mKdV) hierarchies in a single  integrable structure. This relationship between gravitation
	in three-dimensional spacetimes and two-dimensional integrable systems is based on an extension of the recently
	introduced ``soft hairy boundary conditions'' on AdS$_3$, where the chemical potentials are now allowed to depend locally
	on the dynamical fields and their spatial derivatives. The complete integrable structure
	of the Gardner system, i.e., the phase space, the Poisson brackets and the infinite
	number of commuting conserved charges, are directly obtained from the asymptotic
	analysis and the conserved surface integrals in the gravitational theory. 
	These boundary conditions have the particular property that they can also be interpreted
	as being defined in the near horizon region of spacetimes with event horizons. Black hole solutions are then naturally accommodated
	within our boundary conditions, and are described by static configurations
	associated to the corresponding member of the Gardner hierarchy.
	The thermodynamic properties of the black holes in the ensembles
	defined by our boundary conditions are also discussed. Finally, we show that our
	results can be naturally extended to the case of a vanishing cosmological
	constant, and the integrable system turns out to be precisely the same as
	in the case of AdS$_3$.}

\makeatother

\begin{document}
\maketitle \flushbottom

\section{Introduction\label{sec:Introduction}}

The asymptotic structure of General Relativity in 2+1 spacetime dimensions
has been recently shown to be deeply linked with certain special classes
of integrable systems in 1+1 dimensions. For example, a new set of
boundary conditions for Einstein gravity on AdS$_{3}$, labeled by
a nonnegative integer $k$, was proposed in ref. \cite{Perez:2016vqo}.
The dynamics of the gravitational excitations at the boundary are
then described by two independent copies of the $k$-th element of
the Korteweg--de Vries (KdV) hierarchy of nonlinear partial differential
equations, where the standard Brown--Henneaux boundary conditions
\cite{Brown:1986nw} are contained as a particular case ($k=0$).
For a vanishing cosmological constant, the associated integrable system
corresponds to a generalization of the Hirota-Satsuma coupled KdV
system \cite{Hirota:1981wb}, which possesses a BMS$_{3}$ Poisson
structure \cite{Fuentealba:2017omf}.

The key of this relationship between three-dimensional gravity and
two-dimensional integrable systems is the precise way in which the
length and time scales are fixed in the asymptotic region of spacetime
in the gravitational theory. Indeed, the asymptotic values of the
lapse and shift functions in an ADM decomposition of the metric, can
be chosen to depend explicitly on the dynamical fields in a very tight
form, compatible with the action principle.

In the case of the KdV hierarchy analyzed in ref. \cite{Perez:2016vqo},
the fall-off of the metric coincides with the one of Brown and Henneaux
when one includes the most general form of the chemical potentials
\cite{Henneaux:2013dra,Bunster:2014mua}. However, the lapse and shift
functions, and consequently the boundary metric of AdS$_{3}$, are
no longer kept fixed at infinity, because they are now allowed to
explicitly depend on the Virasoro currents. As a consequence of this
particular choice of chemical potentials, Einstein equations in the
asymptotic region precisely reduce to two independent left and right
copies of the $k$-th element of the KdV hierarchy. Furthermore, the
complete integrable structure of the KdV system, i.e., the phase space,
the Poisson brackets and the infinite number of commuting conserved
charges, are directly obtained from the asymptotic analysis and the
conserved surface integrals in the gravitational theory. It is worth
pointing out that there exists a deep relation between the KdV hierarchy
and two-dimensional conformal field theories (CFT$_{2}$), indeed,
the infinite (quantum) commuting KdV charges can be expressed as composite
operators in terms of the stress tensor of a CFT$_{2}$ \cite{Sasaki:1987mm,Eguchi:1989hs,Bazhanov:1994ft}.
This fact has been recently used to describe Generalized Gibbs ensembles
in these theories \cite{deBoer:2016bov,Dymarsky:2018lhf,Maloney:2018hdg,Maloney:2018yrz,Dymarsky:2018iwx,Brehm:2019fyy,Dymarsky:2019etq},
as well as their holographic description \cite{Perez:2016vqo}.

In this paper, we explore the consequences of extending the ``soft
hairy'' boundary conditions on AdS$_{3}$ introduced in refs. \cite{Afshar:2016wfy,Afshar:2016kjj}
by choosing the chemical potentials as appropriate \emph{local} functions
of the fields, and its relation with integrable systems. For this
class of boundary conditions, the fall-off of the metric near infinity
differs from the one of Brown and Henneaux, and they have the particular
property that, by virtue of the topological nature of three-dimensional
General Relativity, they can also be interpreted as being defined
in the near horizon region of spacetimes with event horizons.

The corresponding integrable system is shown to be the Gardner hierarchy
of non-linear partial differential equations, also known as the ``mixed
KdV-mKdV'' hierarchy (see e.g. \cite{Ames1992nonlinear}), whose
first member is given by 
\begin{equation}
\dot{\mathcal{J}}=3a\mathcal{J}\mathcal{J^{\prime}}+3b\mathcal{J}^{2}\mathcal{J^{\prime}}-2\mathcal{J^{\prime\prime\prime}}.\label{eq:gardnereq}
\end{equation}
Here, dots and primes denote derivatives with respect to the temporal
and spatial coordinates respectively, while $a,$ $b$ are arbitrary
constants. Equation \eqref{eq:gardnereq} has the very special property
that simultaneously combines both, KdV and modified KdV (mKdV) equations.
Remarkably, in spite of the deformations generated by the arbitrary
parameters $a$ and $b$, the integrability of this equation and of
the complete hierarchy associated to it, is maintained. Indeed, when
$b=0$ and $a\neq0$, eq. \eqref{eq:gardnereq} precisely reduces
to the KdV equation, while when $a=0$ and $b\neq0$ it coincides
with the mKdV equation\footnote{There is a very deep link between mKdV and KdV equations. If the field
	$\mathcal{J}$ obeys the mKdV equation, then the field\textcolor{red}{{}
	}$\mathcal{L}=b\mathcal{J}^{2}+2\sqrt{b}\mathcal{J}^{\prime}$ will
	precisely obey the KdV equation with $a=1$. This relationship between
	$\mathcal{L}$ and $\mathcal{J}$ is known as the Miura transformation,
	and turns out to be fundamental in order to prove the integrability
	of the KdV and mKdV equations \cite{Miura:1968,Gardner:1968,Kruskal1970,Gardner1974}
	(see also \cite{das1989integrable}).}.\textcolor{red}{{} }From the point of view of the gravitational theory,
there exists a precise choice of chemical potentials, extending the
boundary conditions in \cite{Afshar:2016wfy,Afshar:2016kjj}, such
that Einstein equations reduce to two (left and right) copies of eq.
\eqref{eq:gardnereq}. Moreover, the infinite number of abelian conserved
quantities of the Gardner hierarchy are obtained from the asymptotic
symmetries and the canonical generators in General Relativity using
the Regge-Teitelboim approach \cite{Regge:1974zd}. These results
are generalized to the whole hierarchy, and can also be applied to
the case with a vanishing cosmological constant. We also show that
if the hierarchy is ``extended backwards'' one recovers the soft
hairy boundary conditions of refs. \cite{Afshar:2016wfy,Afshar:2016kjj}
as a particular case.

For a negative cosmological constant, black hole solutions are naturally
accommodated within our boundary conditions, and are described by
static configurations associated to the corresponding member of the
Gardner hierarchy. In particular, BTZ black holes \cite{Banados:1992wn,Banados:1992gq}
are described by constant values of the left and right fields $\mathcal{J}_{\pm}$.
Static solutions with nonconstants $\mathcal{J}_{\pm}$ can also be
found, and describe black holes carrying nontrivial conserved charges.
The thermodynamic properties of the black holes in the ensembles defined
by our boundary conditions will be also discussed.

\emph{Note added:} After this work was completed we received the preprint
\cite{Grumiller:2019tyl}, where the authors also considered the KdV hierarchy
in the context of generalizations of the soft hairy boundary conditions
introduced in refs. \cite{Afshar:2016wfy,Afshar:2016kjj}. In our
notation this corresponds to the particular case $b=0$. However,
they mostly focus on the description of the boundary actions associated
with this integrable system, and its interpretation from a near horizon
perspective.

\section{Asymptotic behavior of the gravitational field\label{sec:Asymp}}

In this section we describe the asymptotic behavior (fall-off) of
the gravitational field without specifying yet what is fixed at the
boundary of spacetime, i.e., without imposing at this step of the
analysis a precise boundary condition. For the purpose of simplicity
and clarity in the presentation, we mostly work in the Chern--Simons
formulation of three-dimensional Einstein gravity with a negative
cosmological constant \cite{Achucarro:1987vz,Witten:1988hc}.

The action for General Relativity on AdS$_{3}$ can be written as
a Chern--Simons action for the gauge group $SL(2,\mathbb{R})\times SL(2,\mathbb{R})$

\[
I=I_{CS}\left[A^{+}\right]-I_{CS}\left[A^{-}\right],
\]
where

\begin{equation}
I_{CS}\left[A\right]=\frac{\kappa}{4\pi}\int\left\langle AdA+\frac{2}{3}A^{3}\right\rangle .\label{eq:Ics}
\end{equation}
Here, the gauge connections $A^{\pm}$ are 1-forms valued on the $sl(2,\mathbb{R})$
algebra, and are related to the vielbein $e$ and spin connection
$\omega$ through $A^{\pm}=\omega\pm e\ell^{-1}$. The level in \eqref{eq:Ics}
is given by $\kappa=\ell/4G$, where $\ell$ is the AdS radius and
$G$ is the Newton constant, while the bilinear form $\left\langle ,\right\rangle $
is defined by the trace in the fundamental representation of $sl(2,\mathbb{R})$.
The $sl(2,\mathbb{R})$ generators $L_{n}$ with $n=0,\pm1$, obey
the commutation relations $\left[L_{n},L_{m}\right]=\left(n-m\right)L_{n+m}$,
with non-vanishing components of the bilinear form given by $\left\langle L_{1}L_{-1}\right\rangle =-1$,
and $\left\langle L_{0}^{2}\right\rangle =1/2$.

\subsection{Asymptotic form of the gauge field\label{subsec:Asymptotic-form-of}}

In order to describe the fall-off of the gauge connection, we closely
follow the analysis in refs. \cite{Afshar:2016wfy,Afshar:2016kjj}.
We will assume that the gauge fields in the asymptotic region take
the form

\begin{equation}
A^{\pm}=b_{\pm}^{-1}\left(d+\mathfrak{a}^{\pm}\right)b_{\pm},\label{eq:gaugetransform}
\end{equation}
where $b_{\pm}$ are gauge group elements depending only on the radial
coordinate, and $\mathfrak{a}^{\pm}=\mathfrak{a}_{t}^{\pm}dt+\mathfrak{a}_{\phi}^{\pm}d\phi$
correspond to auxiliary connections that only depend on the temporal
and angular coordinates $t$ and $\phi$, respectively \cite{Coussaert:1995zp}.
It is worth pointing out that, by virtue of the lack of local propagating
degrees of freedom of three-dimensional Einstein gravity, all the
relevant physical information necessary for the asymptotic analysis
is completely encoded in the auxiliary connections $\mathfrak{a}^{\pm}$,
independently of the precise choice of $b_{\pm}$. We will not specify
any particular $b_{\pm}$ until section \ref{sec:metric}, where the
metric formulation will be discussed.

The auxiliary connections $\mathfrak{a}^{\pm}$ are chosen to be given
by

\begin{equation}
\mathfrak{a}^{\pm}=L_{0}\left(\zeta^{\pm}dt\pm\mathcal{J}^{\pm}d\phi\right).\label{eq:diagonalg}
\end{equation}
Here, $\mathfrak{a}^{\pm}$ are diagonal matrices in the fundamental
representation of $sl(2,\mathbb{R})$, and by that reason we say that
the connections $\mathfrak{a}^{\pm}$ in eq. \eqref{eq:diagonalg}
are written in the ``diagonal gauge''. The fields $\zeta^{\pm}$
are defined along the temporal components of $\mathfrak{a}^{\pm}$,
and consequently they correspond to Lagrange multipliers (chemical
potentials). On the other hand, $\mathcal{J}^{\pm}$ belong to the
spatial components of the gauge connections and therefore they are
identified as the dynamical fields.

The field equations are determined by the vanishing of the field strength
$F^{\pm}=dA^{\pm}+A^{\pm2}=0$, and take the form

\begin{equation}
\dot{\mathcal{J}_{\pm}}=\pm\zeta_{\pm}^{\prime}.\label{eq:fieldeq}
\end{equation}
Note that there are no time derivatives associated to the fields $\zeta_{\pm}$,
which is consistent with the fact that they are chemical potentials.

It is important to emphasize that, as was pointed out in refs. \cite{Afshar:2016wfy,Afshar:2016kjj},
one can also interpret the connections in \eqref{eq:gaugetransform}
and \eqref{eq:diagonalg} as describing the behavior of the fields
in a region near a horizon. Indeed, the reconstructed metric can always
be expressed (in a co-rotating frame) as the direct product of the
two-dimensional Rindler metric times $S^{1}$, together with appropriate
deviations from it. Furthermore, the charges are independent of the
gauge group element $b_{\pm}\left(r\right)$, and consequently they
do not depend on the precise value of the radial coordinate where
they are evaluated. In this sense, one can also interpret the present
analysis as describing ``near horizon boundary conditions''. This
near horizon interpretation can also be extended to higher spacetime
dimensions \cite{SHHD}.

\subsection{Consistency with the action principle and canonical generators}

A fundamental requirement that must be guaranteed is the consistency
of the asymptotic form of the gauge connections in eqs. \eqref{eq:gaugetransform}
and \eqref{eq:diagonalg} with the action principle. In order to analyze
the physical consequences of this requirement, we will use the Regge-Teitelboim
approach \cite{Regge:1974zd} in the canonical formulation of Chern--Simons
theory.

The canonical action can be written as

\[
I_{can}\left[A^{\pm}\right]=-\frac{\kappa}{4\pi}\int dtd^{2}x\varepsilon^{ij}\left\langle A_{i}^{\pm}\dot{A_{j}^{\pm}}-A_{t}^{\pm}F_{ij}^{\pm}\right\rangle +B_{\infty}^{\pm},
\]
where $B_{\infty}^{\pm}$ are boundary terms needed to ensure that
the action principle attains an extremum. Their variations are then
given by

\[
\delta B_{\infty}^{\pm}=-\frac{\kappa}{2\pi}\int dtd\phi\left\langle A_{t}^{\pm}\delta A_{\phi}^{\pm}\right\rangle ,
\]
so that, when one evaluates them for the asymptotic form of the connections
\eqref{eq:gaugetransform} and \eqref{eq:diagonalg} yields 
\begin{equation}
\delta B_{\infty}^{\pm}=\mp\frac{\kappa}{4\pi}\int dtd\phi\zeta_{\pm}\delta\mathcal{J_{\pm}}.\label{eq:deltaB}
\end{equation}

Consistency of the analysis requires that one must be able to ``take
the delta outside'' in the variation of the boundary terms \eqref{eq:deltaB},
i.e., they must be integrable in the functional space\footnote{The requirement of integrability of the boundary terms can relaxed
	in the presence of ingoing or outgoing radiation, where their lack
	of integrability precisely gives the rate of change of the charges
	in time \cite{Barnich:2011mi,Bunster:2018yjr,Bunster:2019mup}. In
	the present case this possibility is not at hand, because General
	Relativity in 2+1-dimensions does not have local propagating degrees
	of freedom.}. This can only be achieved provided one specifies a precise boundary
condition, which turns out to be equivalent to specify what fields
are fixed, i.e., without functional variation, at the boundary of
spacetime. Following ref. \cite{Perez:2016vqo}, a generic possible
choice of boundary conditions is to assume that the chemical potentials
$\zeta_{\pm}$ depend on the dynamical fields $\mathcal{J_{\pm}}$
through

\begin{equation}
\zeta_{\pm}=\frac{4\pi}{\kappa}\frac{\delta H^{\pm}}{\delta\mathcal{J_{\pm}}},\label{eq:zeta}
\end{equation}
where $H^{\pm}=\int d\phi\mathcal{H}^{\pm}[\mathcal{J_{\pm}},\mathcal{J}_{\pm}^{\prime},\mathcal{J_{\pm}}^{\prime\prime},\dots]$
are functionals depending \emph{locally} on the fields $\mathcal{J_{\pm}}$
and their spatial derivatives. Here we also assume that the left and
right sectors are decoupled. With the particular choice of boundary
conditions given by \eqref{eq:zeta}, the delta in \eqref{eq:deltaB}
can be immediately taken outside, and consequently the boundary terms
necessary to improve the canonical action take the form

\[
B^{\pm}=\mp\int dtH^{\pm}.
\]
An immediate consequence of this choice is that the total energy of
the system, defined as the on-shell value of the generator of translations
in time, can be directly written in terms of the ``Hamiltonians''
$H^{\pm}$ as
\begin{equation}
E=H^{+}+H^{-}.\label{eq:Energia}
\end{equation}

The conserved charges associated to the asymptotic symmetries are
also sensitive to the choice of boundary conditions, but in a more
subtle way. The form of the connections $\mathfrak{a}^{\pm}$ in eq.
\eqref{eq:diagonalg} is preserved under gauge transformations $\delta\mathfrak{a}^{\pm}=d\lambda^{\pm}+\left[\mathfrak{a}^{\pm},\lambda^{\pm}\right]$,
with gauge parameters $\lambda^{\pm}$ given by\footnote{Extra terms along $L_{1}$ and $L_{-1}$ might also be added, however
	they are pure gauge.} 
\[
\lambda^{\pm}=\eta_{\pm}L_{0},
\]
provided the fields $\mathcal{J}^{\pm}$ and the chemical potentials
$\zeta^{\pm}$ transform as

\begin{equation}
\delta\mathcal{J_{\pm}}=\pm\eta_{\pm}^{\prime},\label{eq:deltaJ}
\end{equation}
\begin{equation}
\delta\zeta_{\pm}=\dot{\eta}_{\pm}.\label{eq:deltaz}
\end{equation}

Following the Regge-Teitelboim approach \cite{Regge:1974zd}, the
variation of the conserved charges reads

\begin{equation}
\delta Q^{\pm}\left[\eta_{\pm}\right]=\pm\frac{\kappa}{4\pi}\oint d\phi\eta_{\pm}\delta\mathcal{J_{\pm}}.\label{eq:deltaQ}
\end{equation}
Here, the parameters $\eta^{\pm}$ are not arbitrary, because now
the chemical potentials $\zeta^{\pm}$ depend on the fields $\mathcal{J}^{\pm}$
and their spatial derivatives. Hence, one must use the chain rule
in the variation at the left hand-side of eq. \eqref{eq:deltaz},
giving the following first order differential equations in time for
$\eta^{\pm}$

\begin{equation}
\dot{\eta}_{\pm}=\pm\frac{4\pi}{\kappa}\frac{\delta}{\delta\mathcal{J_{\pm}}}\int d\phi\frac{\delta H^{\pm}}{\delta\mathcal{J_{\pm}}}\partial_{\phi}\eta_{\pm}.\label{eq:etapunto}
\end{equation}
Generically, these differential equations will depend explicitly on
$\mathcal{J_{\pm}}$, and consequently finding explicit solutions
to them becomes a very hard task. However, this can be achieved for
certain special choices of boundary conditions, which are related
to the integrable hierarchy associated to the Gardner equation. This
will be discussed in detail in the next section.

\section{Different choices of boundary conditions in the diagonal gauge\label{sec:Different-choices-of}}

\subsection{Soft hairy boundary conditions\label{sec:softhairy}}

A simple choice of boundary conditions corresponds to fix the chemical
potentials $\zeta_{\pm}$ at boundary, such that they are arbitrary
functions without variation, i.e., $\delta\zeta_{\pm}=0$. This possibility
was analyzed in detail in refs. \cite{Afshar:2016wfy,Afshar:2016kjj},
and it was termed ``soft hairy boundary conditions''. The asymptotic
symmetry algebra is then spanned by the generators

\[
Q_{\text{soft hairy}}^{\pm}\left[\eta_{\pm}\right]=\pm\frac{\kappa}{4\pi}\oint d\phi\eta_{\pm}\mathcal{J_{\pm}},
\]
where, by the consistency with time evolution, the parameters $\eta_{\pm}$
are arbitrary functions without variation ($\delta\eta_{\pm}=0$)
. The global charges are then characterized by $\mathcal{J_{\pm}}$,
which obey the following Poisson brackets

\begin{equation}
\left\{ \mathcal{J_{\pm}}\left(\phi\right),\mathcal{J_{\pm}}\left(\bar{\phi}\right)\right\} =\pm\frac{4\pi}{\kappa}\partial_{\phi}\delta\left(\phi-\bar{\phi}\right).\label{eq:Poissonbracket}
\end{equation}
The asymptotic symmetry algebra is then given by two copies of $\hat{u}\left(1\right)$
current algebras.

In a co-rotating frame, $\zeta_{+}=\zeta_{-}=const.$, the generator
of time evolution is identified with the sum of the left and right
zero modes of $\mathcal{J_{\pm}}$, that commute with all the members
of the algebra. In this sense, one can say that the higher modes describe
``soft hair excitations'' in the sense of Hawking, Perry and Strominger
\cite{Hawking:2016msc,Hawking:2016sgy}, because they do not change
the energy of the gravitational configuration. Besides, in this frame,
it is possible to find solutions with non-extremal horizons which
are diffeomorphic to BTZ black holes, and that are endowed with not
trivial soft hair charges. These solutions were called ``black flowers''\footnote{A different type of black flower solution was found in ref. \cite{Barnich:2015dvt}
	in the context of new massive gravity \cite{Bergshoeff:2009hq,Bergshoeff:2009aq}.} \cite{Afshar:2016wfy,Afshar:2016kjj}.

\subsection{Gardner equation\label{sec:mixedmkdv}}

A different choice of boundary conditions that makes contact with
the Gardner (mixed KdV-mKdV) equation is $\zeta^{\pm}=\frac{3}{2}a\mathcal{J_{\pm}}^{2}+b\mathcal{J_{\pm}}^{3}-2\mathcal{J_{\pm}^{\prime\prime}}$,
which according to eq. \eqref{eq:zeta} corresponds to use the following
Hamiltonians
\begin{equation}
H_{\left(1\right)}^{\pm}=\frac{\kappa}{4\pi}\oint d\phi\left(\frac{1}{2}a\mathcal{J}_{\pm}^{3}+\frac{1}{4}b\mathcal{J}_{\pm}^{4}+\mathcal{J}_{\pm}'^{2}\right).\label{eq:H(1)}
\end{equation}
With this particular choice of chemical potentials, Einstein equations
\eqref{eq:fieldeq} precisely reduce to two independent left and right
copies of the Gardner equation \eqref{eq:gardnereq}, that read
\begin{equation}
\dot{\mathcal{J}_{\pm}}=\pm\left(3a\mathcal{J_{\pm}}\mathcal{J_{\pm}^{\prime}}+3b\mathcal{J_{\pm}}^{2}\mathcal{J_{\pm}^{\prime}}-2\mathcal{J_{\pm}^{\prime\prime\prime}}\right).\label{eq:mixedkdvmkdv}
\end{equation}
Equations \eqref{eq:etapunto}, describing the time evolution of the
gauge parameters $\eta^{\pm}$, now take the form 
\begin{equation}
\dot{\eta}_{\pm}=\pm\left(3a\mathcal{J_{\pm}}\eta_{\pm}^{\prime}+3b\mathcal{J_{\pm}}^{2}\eta_{\pm}^{\prime}-2\eta_{\pm}^{\prime\prime\prime}\right),\label{eq:etapunto2}
\end{equation}
where the explicit dependence on $\mathcal{J_{\pm}}$ is manifest.
These equations are linear in $\eta_{\pm}$, and by virtue of the
integrability of the system, it is possible to find their general
solutions under the assumption that they depend locally on $\mathcal{J_{\pm}}$
and their spatial derivatives. Indeed, the gauge parameters $\eta_{\pm}$
obeying eq. \eqref{eq:etapunto2}, are expressed as a linear combination
of ``generalized Gelfand-Dikii polynomials'' $R_{\left(n\right)}^{\pm}$,
i.e.,
\begin{equation}
\eta_{\pm}=\frac{4\pi}{\kappa}\sum_{n\geq0}\alpha_{\left(n\right)}^{\pm}R_{\left(n\right)}^{\pm},\label{eq:etasol}
\end{equation}
where $\alpha_{\left(n\right)}^{\pm}$ are arbitrary constants, and
the polynomials $R_{\left(n\right)}^{\pm}$ are defined through the
following recursion relation
\begin{equation}
\partial_{\phi}R_{(n+1)}^{\pm}=\mathcal{D}_{\phi}R_{(n)}^{\pm},\label{eq:recurrenciarel}
\end{equation}
where $\mathcal{D}_{\phi}$ is a non--local operator given by

\begin{equation}
\mathcal{D}_{\phi}:=a\left(\partial_{\phi}\mathcal{J_{\pm}}+2\mathcal{J_{\pm}}\partial_{\phi}\right)+2b\partial_{\phi}\left(\mathcal{J_{\pm}}\partial_{\phi}^{-1}\left(\mathcal{J_{\pm}}\partial_{\phi}\right)\right)-2\partial_{\phi}^{3}.\label{eq:Dcaligrafico}
\end{equation}
The polynomials $R_{\left(n\right)}^{\pm}$ can be expressed as the
``gradient'' of the Hamiltonians $H_{\left(n\right)}^{\pm}$ of
the integrable system, i.e., 
\begin{equation}
R_{\left(n\right)}^{\pm}=\frac{\delta H_{\left(n\right)}^{\pm}}{\delta\mathcal{J_{\pm}}}.\label{eq:RHJ}
\end{equation}
The first generalized Gelfand-Dikii polynomials together with their
corresponding Hamiltonians are explicitly displayed in appendix \ref{sec: GelfandDikii}.

If one replaces the general solution of eqs. \eqref{eq:etapunto2},
given by \eqref{eq:etasol}, into the expression for the variation
of the canonical generators \eqref{eq:deltaQ}, one can take immediately
the delta outside, and write the conserved charges as

\begin{equation}
Q^{\pm}=\pm\sum_{n\geq0}\alpha_{\left(n\right)}^{\pm}H_{\left(n\right)}^{\pm},\label{eq:QHn}
\end{equation}
where the Hamiltonians $H_{\left(n\right)}^{\pm}$ are in involution
with respect to the Poisson brackets \eqref{eq:Poissonbracket},
\[
\left\{ H_{\left(n\right)}^{\pm},H_{\left(m\right)}^{\pm}\right\} =0.
\]

The presence of the operator $\mathcal{D}_{\phi}$ in the recurrence
relation for the generalized Gelfand-Dikii polynomials \eqref{eq:recurrenciarel},
is related to the fact that this integrable system is bi--Hamiltonian.
Consequently, it is possible to define a second Poisson bracket between
the dynamical fields
\begin{equation}
\left\{ \mathcal{J_{\pm}}\left(\phi\right),\mathcal{J_{\pm}}\left(\bar{\phi}\right)\right\} _{2}:=\pm\frac{4\pi}{\kappa}\mathcal{D}_{\phi}\delta\left(\phi-\bar{\phi}\right),\label{eq:Poissonbracket2}
\end{equation}
so that the Gardner equations can be written as 
\[
\dot{\mathcal{J}_{\pm}}=\left\{ \mathcal{J_{\pm}},H_{\left(1\right)}^{\pm}\right\} =\left\{ \mathcal{J_{\pm}},H_{\left(0\right)}^{\pm}\right\} _{2},
\]
where the Hamiltonians $H_{\left(1\right)}^{\pm}$ are given by eq.
\eqref{eq:H(1)}, while the Hamiltonians $H_{\left(0\right)}^{\pm}$
take the form
\begin{equation}
H_{(0)}^{\pm}=\frac{\kappa}{4\pi}\oint d\phi\left(\frac{1}{2}\mathcal{J}_{\pm}^{2}\right).\label{eq:H0}
\end{equation}

It is worth pointing out that only the first Poisson bracket structure,
given by eq. \eqref{eq:Poissonbracket}, is obtained from the gravitational
theory using the Dirac method for constrained systems. It is not clear
how to obtain the second Poisson structure \eqref{eq:Poissonbracket2}
directly from the canonical structure of General Relativity.

As it was discussed in the introduction, the (left/right) Gardner
equation in \eqref{eq:mixedkdvmkdv} has the special property that
when $a=0$ and $b\neq0$, the equation and its whole integrable structure,
precisely reduce to those of the mKdV integrable system. Conversely,
when $b=0$ and $a\neq0$, the integrable system reduces to KdV. In
this case, the operator $\mathcal{D}_{\phi}$ in \eqref{eq:Dcaligrafico}
becomes a local differential operator, and the second Poisson structure
associated to it through eq. \eqref{eq:Poissonbracket2} corresponds
to two copies of the Virasoro algebra with left and right central
charges given by $c^{\pm}=3\ell/\left(Ga^{2}\right)$. Note that these
central charges do not coincide with the ones of Brown and Henneaux.
This is not surprising because, as we say before, this Poisson structure
does not come directly from the canonical structure of the gravitational
theory.

The Gardner equation does not have additional symmetries besides the
ones generated by the infinite number of commuting Hamiltonians described
above. In particular, it is not invariant under Galilean boosts unless
$b=0$ (KdV case). Nevertheless, it is possible to show that there
exists a particular Galilean boost with parameter $\omega=3a^{2}/\left(4b\right)$,
together with a shift in $\mathcal{J_{\pm}}$ given by $\mathcal{J_{\pm}}\rightarrow\mathcal{J_{\pm}}-a/\left(2b\right)$,
such that the Gardner equation reduces to the mKdV equation. However,
the conserved charges are not mapped into each other, which is a direct
consequence of the fact that this transformation does not map the
higher members of the Gardner and mKdV hierarchies.

In the next section we will show how to extend the previous results
in order to incorporate an arbitrary member of the Gardner hierarchy.

\subsection{Extension to the Gardner hierarchy\label{sec:hierarchy}}

It is possible to find precise boundary conditions for General Relativity
on AdS$_{3}$, such that the dynamics of the boundary degrees of freedom
are described by the $k$-th member of the Gardner hierarchy. This
is achieved by choosing the chemical potentials $\zeta_{\pm}$ as
follows 
\begin{equation}
\zeta_{\pm}=\frac{4\pi}{\kappa}R_{\left(k\right)}^{\pm},\label{eq:zeta2}
\end{equation}
where $R_{\left(k\right)}^{\pm}$ are the $k$-th generalized Gelfand-Dikii
polynomials. The functionals $H^{\pm}$ in eq. \eqref{eq:zeta}, are
identified with the $k$-th Hamiltonian of the hierarchy defined by
eq. \eqref{eq:RHJ}. The particular case $k=1$ corresponds to the
one developed in the previous section.

With the boundary condition specified by eq. \eqref{eq:zeta2}, Einstein
equations in \eqref{eq:fieldeq} take the form 
\begin{equation}
\dot{\mathcal{J}_{\pm}}=\pm\frac{4\pi}{\kappa}\partial_{\phi}R_{\left(k\right)}^{\pm},\label{eq:JdR}
\end{equation}
which coincide with two (left/right) copies of the $k$-th element
of the Gardner hierarchy. By virtue of the bi-Hamiltonian character
of the system, eq. \eqref{eq:JdR} can also be written as
\begin{equation}
\dot{\mathcal{J}_{\pm}}=\left\{ \mathcal{J_{\pm}},H_{\left(k\right)}^{\pm}\right\} =\left\{ \mathcal{J_{\pm}},H_{\left(k-1\right)}^{\pm}\right\} _{2}.\label{eq:jpuntobracket}
\end{equation}

The equations that describe the time evolution of the gauge parameters
$\eta_{\pm}$ take the form of eq. \eqref{eq:etapunto}, but with
$H^{\pm}\rightarrow H_{\left(k\right)}^{\pm}$, whose general solution
that depends locally on $\mathcal{J_{\pm}}$ and their spatial derivatives
coincides with eq. \eqref{eq:etasol}. Consequently, the global charges
integrate as \eqref{eq:QHn}, which as it was discussed in the previous
section, commute among them with both Poisson structures.

\subsubsection*{\emph{Lifshitz scaling.}}

The members of the Gardner hierarchy are not invariant under Lifshitz
scaling. However, in the particular cases when they belong to KdV
or mKdV hierarchies, the anisotropic scaling, with dynamical exponent
$z=2k+1$, is restored.

\emph{Case 1: $a=0$ (mKdV)}

When $a=0$, the $k$-th member of the hierarchy is invariant under

\begin{equation}
t\rightarrow\lambda^{z}t\quad,\quad\phi\rightarrow\lambda\phi\quad,\quad\mathcal{J_{\pm}}\rightarrow\lambda^{-1}\mathcal{J_{\pm}}.\label{eq:LifshitzmKdV}
\end{equation}

\emph{Case 2: $b=0$ (KdV)}

When $b=0$, the $k$-th member of the hierarchy is invariant under

\begin{equation}
t\rightarrow\lambda^{z}t\quad,\quad\phi\rightarrow\lambda\phi\quad,\quad\mathcal{J_{\pm}}\rightarrow\lambda^{-2}\mathcal{J_{\pm}}.\label{eq:LifshitzKdV}
\end{equation}

Note that the dynamical exponent is the same in both cases, but the
fields $\mathcal{J_{\pm}}$ scale in different ways.

\subsubsection*{\emph{Extending the hierarchy backwards.}}

The first nonlinear members of the Gardner hierarchy correspond to
the case $k=1$, and are given by eq. \eqref{eq:mixedkdvmkdv}. However,
if one uses the Hamiltonians $H_{\left(0\right)}^{\pm}$ defined in
eq. \eqref{eq:H0} together with the first Poisson bracket structure
\eqref{eq:Poissonbracket}, one obtains a linear equation describing
left and right chiral movers 
\[
\dot{\mathcal{J_{\pm}}}=\pm\mathcal{J_{\pm}^{\prime}},
\]
which can be considered as the member with $k=0$ of the hierarchy.

From the point of view of the gravitational theory, it is useful to
extend the hierarchy an additional step backwards by using the recursion
relation \eqref{eq:recurrenciarel} in the opposite direction. Hence,
one obtains a new Hamiltonian for each copy of the form
\begin{equation}
H_{\left(-1\right)}^{\pm}=\frac{\kappa}{4\pi}\oint d\phi\left(a^{-1}\mathcal{J_{\pm}}\right).\label{eq:Hm1}
\end{equation}
These Hamiltonians can only be defined when $a\neq0$, and their corresponding
generalized Gelfand-Dikii polynomials $R_{\left(-1\right)}^{\pm}=\kappa/\left(4\pi a\right)$
can be used as a seed that generates the whole hierarchy through the
recursion relation.

Using the Hamiltonians $H_{\left(-1\right)}^{\pm}$ in \eqref{eq:Hm1},
together with the first Poisson bracket structure \eqref{eq:Poissonbracket},
the soft hairy boundary conditions of refs. \cite{Afshar:2016wfy,Afshar:2016kjj},
reviewed in section \ref{sec:softhairy}, are recovered in a co-rotating
frame. In this case, the chemical potentials $\zeta_{\pm}$ are constants
and given by
\[
\zeta_{\pm}=\frac{4\pi}{\kappa}R_{\left(-1\right)}^{\pm}=a^{-1},
\]
while the equations of motion become $\dot{\mathcal{J_{\pm}}}=0$.
In this sense, one can consider the soft hairy boundary conditions
as being part of the hierarchy, and besides as the first member of
it.

\section{Metric formulation\label{sec:metric}}

In this section, we provide a metric description of the results previously
obtained in the context of the Chern--Simons formulation of General
Relativity on AdS$_{3}$. As was discussed in sec. \ref{subsec:Asymptotic-form-of},
the boundary conditions that describe the Gardner hierarchy can be
interpreted as being defined either at infinity or in the near horizon
region. We will analyze these two possible interpretations in the
metric formalism following the lines of ref. \cite{Afshar:2016kjj}.

\subsection{Asymptotic behavior}

The spacetime metric can be directly reconstructed from the Chern--Simons
fields \eqref{eq:gaugetransform}, \eqref{eq:diagonalg}, provided
a particular gauge group element $b_{\pm}\left(r\right)$ is specified.
In order to describe the metric in the asymptotic region, it is useful
to choose 
\[
b_{\pm}\left(r\right)=\exp\left[\pm\frac{1}{2}\log\left(\frac{2r}{\ell}\right)\left(L_{1}-L_{-1}\right)\right].
\]
The expansion of the metric for $r\rightarrow\infty$ then reads

\begin{eqnarray}
g_{tt} & = & -\zeta_{+}\zeta_{-}r^{2}+\frac{\ell^{2}}{4}\left(\zeta_{+}^{2}+\zeta_{-}^{2}\right)+\mathcal{O}\left(r^{-1}\right),\nonumber \\
g_{tr} & = & \mathcal{O}\left(r^{-2}\right),\nonumber \\
g_{t\phi} & = & \left(\zeta_{+}\mathcal{J}_{-}-\zeta_{-}\mathcal{J}_{+}\right)\frac{r^{2}}{2}+\frac{\ell^{2}}{4}\left(\zeta_{+}\mathcal{J}_{+}-\zeta_{-}\mathcal{J}_{-}\right)+\mathcal{O}\left(r^{-1}\right),\label{eq:AsymptoticMetric}\\
g_{rr} & = & \frac{\ell^{2}}{r^{2}}+\mathcal{O}\left(r^{-5}\right),\nonumber \\
g_{r\phi} & = & \mathcal{O}\left(r^{-2}\right),\nonumber \\
g_{\phi\phi} & = & \mathcal{J_{+}}\mathcal{J_{-}}r^{2}+\frac{\ell^{2}}{4}\left(\mathcal{J}_{+}^{2}+\mathcal{J}_{-}^{2}\right)+\mathcal{O}\left(r^{-1}\right).\nonumber 
\end{eqnarray}

As it was explained in sec. \ref{sec:hierarchy}, in order to implement
the boundary conditions associated to the Gardner hierarchy we must
choose the chemical potentials $\zeta_{\pm}$ according to eq. \eqref{eq:zeta2},
i.e.,
\[
\zeta_{\pm}=\frac{4\pi}{\kappa}R_{\left(k\right)}^{\pm}.
\]
The differential equations associated to the (left/right) $k$-th
element of the hierarchy are precisely recovered if one imposes that
the metric in eq. \eqref{eq:AsymptoticMetric} obeys Einstein equations
with a negative cosmological constant in the asymptotic region of
spacetime.

The fall-off in \eqref{eq:AsymptoticMetric} is preserved under the
asymptotic symmetries generated by the following asymptotic Killing vectors

\begin{eqnarray}
\xi^{t} & = & \frac{\eta^{+}\mathcal{J}_{-}+\eta^{-}\mathcal{J}_{+}}{\zeta_{+}\mathcal{J}_{-}+\zeta_{-}\mathcal{J}_{+}}+\mathcal{O}\left(\frac{1}{r^{3}}\right),\nonumber \\
\xi^{r} & = & \mathcal{O}\left(\frac{1}{r^{2}}\right),\\
\xi^{\phi} & = & \frac{\eta^{+}\zeta_{-}-\eta^{-}\zeta_{+}}{\zeta_{+}\mathcal{J}_{-}+\zeta_{-}\mathcal{J}_{+}}+\mathcal{O}\left(\frac{1}{r^{3}}\right).\nonumber 
\end{eqnarray}
The conserved charges can be directly computed using the Regge-Teitelboim
approach \cite{Regge:1974zd}, and as expected coincide with the expression
in eq. \eqref{eq:deltaQ} obtained using the Chern--Simons formulation.

Note that the boundary metric

\[
d\bar{s}^{2}=r^{2}\left(-\zeta_{+}\zeta_{-}dt^{2}+\left(\zeta_{+}\mathcal{J}_{-}-\zeta_{-}\mathcal{J}_{+}\right)dtd\phi+\mathcal{J_{+}}\mathcal{J_{-}}d\phi^{2}\right),
\]
explicitly depends on the dynamical fields $\mathcal{J_{\pm}}$ and
consequently \emph{it is not fixed} at the boundary of spacetime,
i.e., it has a nontrivial functional variation.

\subsection{Near horizon behavior}

Following ref. \cite{Afshar:2016kjj}, the metric in the near horizon
region can be reconstructed using

\[
b_{\pm}\left(r\right)=\exp\left(\pm\frac{r}{2\ell}\left(L_{1}-L_{-1}\right)\right),
\]
and considering an expansion around $r=0$. The metric then reads

\begin{eqnarray}
g_{tt} & = & \frac{\ell^{2}}{4}\left(\zeta_{+}-\zeta_{-}\right)^{2}-\zeta_{+}\zeta_{-}r^{2}+\mathcal{O}\left(r^{3}\right),\nonumber \\
g_{tr} & = & \mathcal{O}\left(r^{2}\right),\nonumber \\
g_{t\phi} & = & \frac{\ell^{2}}{4}\left(\mathcal{J}_{+}+\mathcal{J}_{-}\right)\left(\zeta_{+}-\zeta_{-}\right)+\left(\zeta_{+}\mathcal{J}_{-}-\zeta_{-}\mathcal{J}_{+}\right)\frac{r^{2}}{2}+\mathcal{O}\left(r^{3}\right),\label{eq:NearHorizonMetric}\\
g_{rr} & = & 1+\mathcal{O}\left(r^{2}\right),\nonumber \\
g_{r\phi} & = & \mathcal{O}\left(r^{2}\right),\nonumber \\
g_{\phi\phi} & = & \frac{\ell^{2}}{4}\left(\mathcal{J}_{+}+\mathcal{J}_{-}\right)^{2}+\mathcal{J}_{+}\mathcal{J}_{-}r^{2}+\mathcal{O}\left(r^{3}\right).\nonumber 
\end{eqnarray}
Again, the chemical potentials $\zeta_{\pm}$ are expressed in terms
of the generalized Gelfand-Dikii polynomials according to eq. \eqref{eq:zeta2}\footnote{Note that with our choice of boundary conditions it is not possible to write the metric \eqref{eq:NearHorizonMetric} in a co-rotating frame ($\zeta_{+}=\zeta_{-}=const.$), because $\zeta_{\pm}$ have a very precise dependence on the fields $\mathcal{J_{\pm}}$, and generically cannot be set to be equal to constants.}.

The behavior of the metric near the horizon is preserved under the
action of the following near horizon Killing vectors

\begin{eqnarray}
\xi^{t} & = & \frac{\eta^{+}\mathcal{J}_{-}+\eta^{-}\mathcal{J}_{+}}{\zeta_{+}\mathcal{J}_{-}+\zeta_{-}\mathcal{J}_{+}}+\mathcal{O}\left(r^{3}\right),\nonumber \\
\xi^{r} & = & \mathcal{O}\left(r^{3}\right),\\
\xi^{\phi} & = & \frac{\eta^{+}\zeta_{-}-\eta^{-}\zeta_{+}}{\zeta_{+}\mathcal{J}_{-}+\zeta_{-}\mathcal{J}_{+}}+\mathcal{O}\left(r^{3}\right).\nonumber 
\end{eqnarray}
The conserved charges can be obtained using the Regge-Teitelboim approach,
and evaluating them at $r=0$. The results coincide with eq. \eqref{eq:deltaQ},
as expected.

\subsection{General solution}

In ref. \cite{Afshar:2016wfy,Afshar:2016kjj}, it was shown that it
is possible to construct the general solution of Einstein equations
obeying the fall-off described in \eqref{eq:AsymptoticMetric}. It
is given by
\begin{equation}
\begin{array}{ccc}
ds^{2} & = & dr^{2}+\frac{\ell^{2}}{4}\cosh^{2}\left(r/\ell\right)\left[\left(\zeta_{+}-\zeta_{-}\right)dt+\left(\mathcal{J}_{+}+\mathcal{J}_{-}\right)d\phi\right]^{2}\\
&  & -\frac{\ell^{2}}{4}\sinh^{2}\left(r/\ell\right)\left[\left(\zeta_{+}+\zeta_{-}\right)dt+\left(\mathcal{J}_{+}-\mathcal{J}_{-}\right)d\phi\right]^{2},
\end{array}\label{eq:metriccompleta-1}
\end{equation}
and satisfies Einstein equations provided that $\mathcal{J}_{\pm}$
obey the differential equations associated to the $k$-th member of
the hierarchy when $\zeta_{\pm}$ is fixed according to eq. \eqref{eq:zeta2}.
In the near horizon region, this solution also obeys the fall-off
in \eqref{eq:NearHorizonMetric}. Note that the metric \eqref{eq:metriccompleta-1}
is diffeomorphic to a BTZ geometry, but as we will show in the next
section, it carries nontrivial charges associated to improper (large)
gauge transformations \cite{Benguria:1977in}, and consequently describes
a different physical state.

It is worth emphasizing that there is a one-to-one map between three--dimensional
geometries described by eq. \eqref{eq:metriccompleta-1}, and solutions
of the members of the Gardner hierarchy. In this sense, we can say
that this integrable system was ``fully geometrized'' in terms of
certain three--dimensional spacetimes which are locally of constant
curvature.

\section{Black holes\label{sec:blackholes}}

\subsection{Regularity conditions and thermodynamics}

Euclidean black holes solutions are obtained by requiring regularity
of the Euclidean geometries associated to the family of metrics in
\eqref{eq:metriccompleta-1}. This fixes the inverse of left and right
temperatures $\beta_{\pm}=T_{\pm}^{-1}$ in terms of the fields $\zeta_{\pm}$
according to
\begin{equation}
\beta_{\pm}=\frac{2\pi}{\zeta_{\pm}}.\label{eq:betamasmenos}
\end{equation}
These conditions can also be obtained by requiring that the holonomy
around the thermal cycle for the gauge connections \eqref{eq:diagonalg}
be trivial.

A direct consequence of eq. \eqref{eq:betamasmenos} is that the chemical
potentials $\zeta_{\pm}$ are now constants, and hence from the field
equations \eqref{eq:fieldeq}, the regular Euclidean solutions are
characterized by $\dot{\mathcal{J_{\pm}}}=0$, i.e., by static solutions
of the members of the Gardner hierarchy. In sum, in order to obtain
an explicit black hole solution, the following equations must be solved

\begin{equation}
\partial_{\phi}R_{\left(k\right)}^{\pm}=0,\label{eq:DR=00003D0}
\end{equation}
restricted to the conditions

\begin{equation}
T_{\pm}=\frac{2}{\kappa}R_{\left(k\right)}^{\pm}.\label{eq:Temperature}
\end{equation}

The Bekenstein--Hawking entropy can be directly obtained from the
near horizon expansion \eqref{eq:NearHorizonMetric}, and gives
\begin{equation}
S=\frac{A}{4G}=\frac{\kappa}{2}\oint d\phi\left(\mathcal{J_{+}}+\mathcal{J_{-}}\right).\label{eq:HawkingBek-Entropy}
\end{equation}
As expected, the first law is automatically fulfilled. Indeed, using
\eqref{eq:Temperature} one obtains

\[
\beta_{+}\delta H_{\left(k\right)}^{+}+\beta_{-}\delta H_{\left(k\right)}^{-}=\oint d\phi\left(\beta_{+}R_{\left(k\right)}^{+}\delta\mathcal{J_{+}}+\beta_{-}R_{\left(k\right)}^{-}\delta\mathcal{J_{-}}\right)=\delta\left[\frac{\kappa}{2}\oint d\phi\left(\mathcal{J_{+}}+\mathcal{J_{-}}\right)\right]=\delta S.
\]
Here, $\beta_{\pm}$ turn out to be the conjugates to the left and
right energies $H_{\left(k\right)}^{\pm}$. The inverse temperature,
conjugate to the energy $E$ in eq. \eqref{eq:Energia}, is expressed
in terms of the left and right temperatures according to $T^{-1}=\frac{1}{2}\left(T_{+}^{-1}+T_{-}^{-1}\right)$.

The previous analysis was performed in a rather abstract form without
using an explicit solution to eq. \eqref{eq:DR=00003D0}, which in
general are very hard to find. A simple solution corresponds to $\mathcal{J_{\pm}}=const.$,
which describes a BTZ configuration. In this case the Hamiltonians
take the form

\[
H_{\left(k\right)}^{\pm}=\sum_{n=k+2}^{2k+2}\alpha_{n}^{\pm}\mathcal{J}_{\pm}^{n},
\]
where $\alpha_{n}^{\pm}$ are constant coefficients which are not
specified in general, but whose values can be determined once the
corresponding Hamiltonians are explicitly computed through the recursion
relation for the generalized Gelfand-Dikii polynomials.

Some simplifications occur when we turn off either $a$ or $b$ (mKdV
and KdV cases), that we discuss next.

\subsection{mKdV case ($a=0$)}

When $a=0$, the metric associated to the black hole solution with
$\mathcal{J_{\pm}}=const.$ can be written as
\begin{equation}
\begin{array}{ccc}
ds^{2} & = & dr^{2}+\frac{\ell^{2}}{4}\cosh^{2}\left(r/\ell\right)\left[4\pi^{2}\left(\frac{\pi\kappa}{2\sigma_{\left(k\right)}\left(k+1\right)}\right)^{2k+1}\left(\mathcal{J}_{+}^{2k+1}-\mathcal{J}_{-}^{2k+1}\right)dt+\left(\mathcal{J}_{+}+\mathcal{J}_{-}\right)d\phi\right]^{2}\\
&  & -\frac{\ell^{2}}{4}\sinh^{2}\left(r/\ell\right)\left[4\pi^{2}\left(\frac{\pi\kappa}{2\sigma_{\left(k\right)}\left(k+1\right)}\right)^{2k+1}\left(\mathcal{J}_{+}^{2k+1}+\mathcal{J}_{-}^{2k+1}\right)dt+\left(\mathcal{J}_{+}-\mathcal{J}_{-}\right)d\phi\right]^{2},
\end{array}\label{eq:metriccompleta-1-1-1}
\end{equation}
where the constant $\sigma_{\left(k\right)}$, given by
\[
\sigma_{\left(k\right)}:=\left(\frac{\pi\kappa}{2k+2}\right)^{\frac{k+1}{k+\frac{1}{2}}}\left(\frac{\sqrt{\pi}}{\kappa2^{k-2}b^{k}}\frac{\Gamma\left(k+2\right)}{\Gamma\left(k+\frac{1}{2}\right)}\right)^{\frac{1}{2k+1}},
\]
will play the role of the anisotropic Stefan--Boltzmann constant
of the system. The metric can be written in Schwarzschild--like coordinates
using the following coordinate transformation
\begin{equation}
r=\frac{\ell}{2}\log\left(\frac{4\sqrt{\left(\bar{r}^{2}-\frac{\ell^{2}}{4}\left(\mathcal{J}_{+}-\mathcal{J}_{-}\right)^{2}\right)\left(\bar{r}^{2}-\frac{\ell^{2}}{4}\left(\mathcal{J}_{+}+\mathcal{J}_{-}\right)^{2}\right)}-\ell^{2}\left(\mathcal{J}_{+}^{2}+\mathcal{J}_{-}^{2}\right)+4\bar{r}^{2}}{2\ell^{2}\mathcal{J}_{+}\mathcal{J}_{-}}\right).\label{eq:changeinr}
\end{equation}
It coincides with the metric of a BTZ black hole in a rotating frame,
with outer and inner horizons located at $\bar{r}_{\pm}=\frac{\ell}{2}\left(\mathcal{J}_{+}\pm\mathcal{J}_{-}\right)$.

With the choice $a=0$, the expression for the left and right energies
written in terms of the constants $\mathcal{J}_{\pm}$ becomes simpler
than the one in the general case. Indeed, it can be written in a closed
form as
\begin{equation}
H_{\left(k\right)}^{\pm}=\sigma_{\left(k\right)}^{-z}\left(\frac{\pi\kappa}{\left(z+1\right)}\mathcal{J}_{\pm}\right)^{z+1},\label{eq:HdeJmKdV}
\end{equation}
where $z=2k+1$ is the dynamical exponent of the Lifshitz scale symmetry
\eqref{eq:LifshitzmKdV} of the $k$-th element of the mKdV hierarchy.

Using eqs. \eqref{eq:Temperature} and \eqref{eq:HdeJmKdV}, the left
and right energies $H_{\left(k\right)}^{\pm}$ can be expressed in
terms of the left and right temperatures $T_{\pm}$, acquiring the
form dictated by the Stefan--Boltzmann law for a two--dimensional
system with anisotropic Lifshitz scaling \cite{Gonzalez:2011nz}

\[
H_{\left(k\right)}^{\pm}=\sigma_{\left(k\right)}T_{\pm}^{1+\frac{1}{z}}.
\]

The Bekenstein--Hawking entropy, given by eq. \eqref{eq:HawkingBek-Entropy},
can be expressed in terms of the left and right energies $H_{\left(k\right)}^{\pm}$
as follows
\begin{equation}
S=\left(1+z\right)\sigma_{\left(k\right)}^{\frac{z}{1+z}}\left(\left(H_{\left(k\right)}^{+}\right)^{\frac{1}{z+1}}+\left(H_{\left(k\right)}^{-}\right)^{\frac{1}{z+1}}\right).\label{eq:EntropyEnergy}
\end{equation}
Note that the dependence of the entropy in terms of the left/right
energies is consistent with the Lifshitz scaling of the $k$--th
element of the mKdV hierarchy.

\subsubsection*{\emph{Power partitions and microstate counting}}

The dependence of the entropy in terms of the left/right energies
in eq. \eqref{eq:EntropyEnergy} might be understood from a microscopic
point of view if, following \cite{Melnikov:2018fhb}, we assume that
there exists a two--dimensional field theory with Lifshitz scaling,
defined on a circle, whose dispersion relation for very high energies
takes the form

\begin{equation}
E_{n}^{\pm}=\varepsilon_{\left(z\right)}^{\pm}n^{z},\label{eq:Dispersion}
\end{equation}
where $n$ is a non--negative integer, and $\varepsilon_{\left(z\right)}^{\pm}$
denote the characteristic energy of the left/right modes. The problem
of computing the entropy in the microcanonical ensemble is then equivalent
to compute the power partitions of given integers $N_{\pm}=E^{\pm}/\varepsilon_{\left(z\right)}^{\pm}$.
Here $E^{\pm}$ are the left/right energies given by 
\[
E^{\pm}=\varepsilon_{\left(z\right)}^{\pm}\sum_{i}n_{i}^{z}.
\]
This problem was solved long ago by Hardy and Ramanujan in \cite{hardy1918},
where at the end of their paper they conjecture that the asymptotic
growth of power partitions is given by

\[
p_{z}\left(N_{\pm}\right)\approx\exp\left[\left(1+z\right)\left(\frac{\Gamma\left(1+\frac{1}{z}\right)\zeta\left(1+\frac{1}{z}\right)}{z}\right)^{\frac{z}{1+z}}N_{\pm}^{\frac{1}{1+z}}\right],
\]
result that was proven later by Wright in 1934 \cite{wright1934asymptotic}.

The (left/right) entropies then reads

\begin{equation}
S^{\pm}=\log\left[p_{z}\left(N_{\pm}\right)\right]=\left(1+z\right)\left(\frac{\Gamma\left(1+\frac{1}{z}\right)\zeta\left(1+\frac{1}{z}\right)}{z}\right)^{\frac{z}{1+z}}\left(\frac{E^{\pm}}{\varepsilon_{\left(z\right)}^{\pm}}\right)^{\frac{1}{1+z}}.\label{eq:EntropyWright}
\end{equation}
This expression precisely coincides with the entropy of the black
hole in eq. \eqref{eq:EntropyEnergy}, provided $E^{\pm}=H_{\left(k\right)}^{\pm}$,
and
\[
\varepsilon_{\left(z\right)}^{\pm}=\left(\frac{\Gamma\left(1+\frac{1}{z}\right)\zeta\left(1+\frac{1}{z}\right)}{\sigma_{\left(k\right)}z}\right)^{z}.
\]

Note that AdS spacetime is not contained within the spectrum of our
boundary conditions, and consequently one can naively think that the
anisotropic extension of Cardy formula of refs. \cite{Gonzalez:2011nz,Perez:2016vqo}
cannot be used to reproduce the entropy of the black hole \eqref{eq:EntropyEnergy}.
However, there is a known case \cite{Gonzalez:2011nz}, where the
anisotropic extension of Cardy formula can still be used in spite
of the fact that ground state, given by a gravitational soliton, does
not fit within the boundary conditions that accommodate the Lifshitz
black hole. This approach is based on the use of an anisotropic extension
of modular invariance that relates the Euclidean black hole and its
corresponding soliton, which turn out to be diffeomorphic. It would
be interesting to explore in the future whether this approach could
be applied to the BTZ black holes in the context of our boundary conditions.

\subsection{KdV case ($b=0$)}

When $b=0$, the metric associated to the black hole solution with
$\mathcal{J_{\pm}}=const.$ takes the form

\begin{equation}
\begin{array}{ccc}
ds^{2} & = & dr^{2}+\frac{\ell^{2}}{4}\cosh^{2}\left(r/\ell\right)\left[4\pi^{2}\left(\frac{\pi\kappa}{\bar{\sigma}_{\left(k\right)}\left(k+2\right)}\right)^{k+1}\left(\mathcal{J}_{+}^{k+1}-\mathcal{J}_{-}^{k+1}\right)dt+\left(\mathcal{J}_{+}+\mathcal{J}_{-}\right)d\phi\right]^{2}\\
&  & -\frac{\ell^{2}}{4}\sinh^{2}\left(r/\ell\right)\left[4\pi^{2}\left(\frac{\pi\kappa}{\bar{\sigma}_{\left(k\right)}\left(k+2\right)}\right)^{k+1}\left(\mathcal{J}_{+}^{k+1}+\mathcal{J}_{-}^{k+1}\right)dt+\left(\mathcal{J}_{+}-\mathcal{J}_{-}\right)d\phi\right]^{2},
\end{array}\label{eq:metriccompleta-1-1}
\end{equation}
where

\[
\bar{\sigma}_{\left(k\right)}=\left(\frac{\pi\kappa}{k+2}\right)^{\frac{k+2}{k+1}}\left(\frac{\sqrt{\pi}}{\kappa\left(2a\right)^{k}}\frac{\Gamma\left(k+3\right)}{\Gamma\left(k+\frac{3}{2}\right)}\right)^{\frac{1}{k+1}}.
\]
Using the change of coordinates \eqref{eq:changeinr}, the metric
\eqref{eq:metriccompleta-1-1} can be written in Schwarzschild--like
form, and coincides with the one of a BTZ black hole with outer and
inner horizons located at $\bar{r}_{\pm}=\frac{\ell}{2}\left(\mathcal{J}_{+}\pm\mathcal{J}_{-}\right)$.

The left and right energies $H_{\left(k\right)}^{\pm}$ can then be
expressed in terms of the constants $\mathcal{J}_{\pm}$ according
to
\begin{equation}
H_{\left(k\right)}^{\pm}=\bar{\sigma}_{\left(k\right)}^{-\frac{z+1}{2}}\left(\frac{2\pi\kappa}{z+3}\mathcal{J}_{\pm}\right)^{\frac{z+3}{2}},\label{eq:HdeJKdV}
\end{equation}
where $z=2k+1$ is the dynamical exponent associated to the Lifshitz
symmetry \eqref{eq:LifshitzKdV} of the $k$-th member of the KdV
hierarchy.

The expression for the left/right energies $H_{\left(k\right)}^{\pm}$
in terms of the left/right temperatures $T_{\pm}$, is then given
by

\[
H_{\left(k\right)}^{\pm}=\bar{\sigma}_{\left(k\right)}T_{\pm}^{\frac{z+3}{z+1}}.
\]
In spite of the fact that the $k$-th equation of the KdV hierarchy
is invariant under Lifshitz scaling with dynamical exponent $z$,
the power in the temperature is not the one expected for a two--dimensional
theory with this symmetry. Furthermore, this is inherited to the expression
for the entropy written in terms of the left/right energies $H_{\left(k\right)}^{\pm}$
\[
S=\left(\frac{z+3}{2}\right)\bar{\sigma}_{\left(k\right)}^{\frac{z+1}{z+3}}\left(\left(H_{\left(k\right)}^{+}\right)^{\frac{2}{z+3}}+\left(H_{\left(k\right)}^{-}\right)^{\frac{2}{z+3}}\right),
\]
which is not of the expected form \eqref{eq:EntropyWright}.

Remarkably, if instead of the Hamiltonians $H_{\left(k\right)}^{\pm}$,
one uses the Hamiltonians $H_{\left(2k\right)}^{\pm}$, this naive
incompatibility with the Lifshitz symmetry disappears. Indeed, the
relation between $H_{\left(2k\right)}^{\pm}$ and the left/right temperatures
takes the form

\[
H_{\left(2k\right)}^{\pm}=\bar{\sigma}_{\left(2k\right)}T_{\pm}^{1+\frac{1}{z}},
\]
while the expression for the entropy in terms of the extensive quantities
$H_{\left(2k\right)}^{\pm}$ reads
\[
S=\left(1+z\right)\bar{\sigma}_{\left(2k\right)}^{\frac{z}{1+z}}\left(\left(H_{\left(2k\right)}^{+}\right)^{\frac{1}{z+1}}+\left(H_{\left(2k\right)}^{-}\right)^{\frac{1}{z+1}}\right).
\]
The entropy then takes the expected Hardy--Ramanujan form \eqref{eq:EntropyWright},
with the characteristic energy of the dispersion relation given by
\[
\varepsilon_{\left(z\right)}^{\pm}=\left(\frac{\Gamma\left(1+\frac{1}{z}\right)\zeta\left(1+\frac{1}{z}\right)}{\bar{\sigma}_{\left(2k\right)}z}\right)^{z}.
\]

\subsubsection*{\emph{Black hole with nonconstants $\mathcal{J_{\pm}}$}}

In the particular case when $b=0$ and $k=1$, it is possible to find
explicit nonconstants solutions to eq. \eqref{eq:DR=00003D0} that
characterize a regular Euclidean black hole. These are static solutions
of the left/right KdV equations, which take the form of periodic cnoidal
waves. The solutions are then given by
\begin{equation}
\mathcal{J_{\pm}}=-\frac{8K^{2}\left(m_{\pm}\right)}{3a\pi^{2}}\left(1-2m_{\pm}+3m_{\pm}\text{cn}^{2}\left(\frac{K\left(m_{\pm}\right)}{\pi}\phi|m^{\pm}\right)\right),
\end{equation}
where $m_{\pm}$ are constants in the range $0\leq m_{\pm}<1$, $\text{cn}$
denotes the Jacobi elliptic cosine function, and $K\left(m\right)$
is the complete elliptic integral of the first kind defined as

\begin{equation}
K(m)=\int_{0}^{\frac{\pi}{2}}\frac{d\theta}{\sqrt{1-m\sin^{2}\theta}}.
\end{equation}
The regularity conditions \eqref{eq:Temperature} fix the left and
right temperatures $T_{\pm}$ in terms of the constant $m_{\pm}$
according to
\[
T_{\pm}=\frac{16\left(m_{\pm}^{2}-m_{\pm}+1\right)}{3\pi^{5}a}K\left(m_{\pm}\right)^{4}.
\]

\section{Final remarks \label{sec: remarks}}

We have shown that by imposing appropriate boundary conditions to
General Relativity on AdS$_{3}$, one can describe the boundary dynamics
of the gravitational field by the integrable system corresponding
to the Gardner (mixed KdV-mKdV) hierarchy. These results can be naturally
extended to the case with a vanishing cosmological constant. Indeed,
following ref. \cite{Afshar:2016kjj}, if one chooses an auxiliary
gauge connection of the form

\begin{equation}
\mathfrak{a}=\frac{1}{2}\left(\left(\zeta_{+}+\zeta_{-}\right)L_{0}+\left(\zeta_{+}-\zeta_{-}\right)P_{0}\right)dt+\frac{1}{2}\left(\left(\mathcal{J_{+}}-\mathcal{J_{-}}\right)L_{0}+\left(\mathcal{J_{+}}+\mathcal{J_{-}}\right)P_{0}\right)d\phi,\label{eq:connectionflat}
\end{equation}
the field equations precisely reduce to eq. \eqref{eq:fieldeq}, i.e.,
\[
\dot{\mathcal{J}_{\pm}}=\pm\zeta_{\pm}^{\prime}.
\]
Here $P_{0}$ and $L_{0}$ are the zero modes of the $isl\left(2\right)$
generators, whose non--vanishing brackets are $\left[L_{n},L_{m}\right]=\left(n-m\right)L_{n+m}$
and $\left[L_{n},P_{m}\right]=\left(n-m\right)P_{n+m}$, with $n,m=0,\pm1$.
Furthermore, the expression for the variation of the charges acquire
precisely the form \eqref{eq:deltaQ}, with $\kappa=1/\left(4G\right)$.
Consequently, the complete integrable structure associated to the
Gardner hierarchy is recovered provided we fix the chemical potentials
$\zeta_{\pm}$ according to eq. \eqref{eq:zeta}, i.e., 
\[
\zeta_{\pm}=\frac{4\pi}{\kappa}R_{\left(k\right)}^{\pm}.
\]
Remarkably, the boundary dynamics in the absence of a cosmological
constant, described by the connection \eqref{eq:connectionflat},
is associated to the same integrable system than in the case of a
negative cosmological constant. In this sense, there is a certain
universality in our results.

The integrable systems discussed here possess an infinite number of
conserved charges that commute among them. This opens the possibility
of studying more general thermodynamic ensembles containing not only
the left/right Hamiltonians, but also the additional conserved charges.
This kind of ensembles are called Generalized Gibbs Ensembles and
have been recently studied in the context of the KdV charges \cite{deBoer:2016bov,Perez:2016vqo,Dymarsky:2018lhf,Maloney:2018hdg,Maloney:2018yrz,Dymarsky:2018iwx,Brehm:2019fyy,Dymarsky:2019etq}.
In our analysis, the conserved charges are more general than the ones
of KdV, which are obtained as a particular case ($b=0$). One could
then investigate the behavior of the Generalized Gibbs Ensembles in
this more general setup, as well as their consequences for the corresponding
dual gravitational theory.

One can also investigate the possibility of introducing suitable ``potentials''
for the fields $\mathcal{J_{\pm}}$ to define an appropriate action
principle, for which the conserved charges $H_{\left(k\right)}^{\pm}$
are obtained from the Noether theorem. This could be achieved by performing
a Hamiltonian reduction along the lines of \cite{Coussaert:1995zp,Gonzalez:2018jgp},
in order to find the corresponding action describing the ``boundary
dynamics''.

It would be also interesting to explore the generalization to the
case of three-dimensional higher spin gravity \cite{Blencowe:1988gj,Bergshoeff:1989ns,Vasiliev:1995dn,Henneaux:2010xg,Campoleoni:2010zq},
where one can use the ``diagonal gauge'' introduced in ref. \cite{Grumiller:2016kcp}
and choose the chemical potentials as appropriate functions of the
dynamical fields. One would expect that, for the particular case of
spin $s=3$, the corresponding integrable system would be related
to the Boussinesq hierarchy and its corresponding modified version.
This might be also extended to the case of higher spin gravity with
vanishing cosmological constant, using the fall-off described in \cite{Ammon:2017vwt}.

A new class of deformations of two-dimensional field theories preserving
integrability was introduced in ref. \cite{Smirnov:2016lqw}. In particular,
deformations of the form $T\bar{T}$ \cite{Zamolodchikov:2004ce,Cavaglia:2016oda,Cardy:2018sdv,Aharony:2018vux,Datta:2018thy,Aharony:2018bad},
and $J\bar{T}$ \cite{Guica:2017lia,Chakraborty:2018vja,Aharony:2018ics,Guica:2019vnb},
as well as their holographic descriptions \cite{McGough:2016lol,Giveon:2017nie,Kraus:2018xrn,Bzowski:2018pcy},
have recently received a great deal of attention. It would be interesting
to apply this class of deformations to the integrable systems studied
in this work, and explore their consequences in the context of the
dual gravitational theory through the boundary conditions proposed
here.

\acknowledgments{We thank Oscar Fuentealba, Daniel Grumiller, Pablo Rodr\'iguez and Ricardo Troncoso for useful discussions. The work of EO was partially funded by the PhD grant CONICYT-PCHA/Doctorado Nacional/2016-21161352. This research has been partially supported by Fondecyt grants N$\textsuperscript{\underline{o}}$ 1171162, 1181496, 1181031, and the grant CONICYT PCI/REDES 170052. The Centro de Estudios Cient\'ificos (CECs) is funded by the Chilean Government through the Centers of Excellence Base Financing Program of Conicyt.}

\appendix

\section{Hamiltonians and generalized Gelfand-Dikii polynomials\label{sec: GelfandDikii}}

In this appendix, we explicitly display the first Hamiltonians $H_{\left(k\right)}^{\pm}$
associated to the Gardner integrable system, as well as their corresponding
generalized Gelfand-Dikii polynomials $R_{\left(k\right)}^{\pm}$.

The first Hamiltonians $H_{\left(k\right)}^{\pm}$ are given by

\begin{eqnarray}
H_{\left(-1\right)}^{\pm} & = & \frac{\kappa}{4\pi}\oint d\phi\left(a^{-1}\mathcal{J_{\pm}}\right),\nonumber \\
H_{\left(0\right)}^{\pm} & = & \frac{\kappa}{4\pi}\oint d\phi\left(\frac{1}{2}\mathcal{J}_{\pm}^{2}\right),\nonumber \\
H_{\left(1\right)}^{\pm} & = & \frac{\kappa}{4\pi}\oint d\phi\left(\frac{1}{2}a\mathcal{J}_{\pm}^{3}+\frac{1}{4}b\mathcal{J}_{\pm}^{4}+\mathcal{J}_{\pm}^{\prime2}\right),\label{eq:k-thH}\\
H_{\left(2\right)}^{\pm} & = & \frac{\kappa}{4\pi}\oint d\phi\left(\frac{5}{8}a^{2}\mathcal{J}_{\pm}^{4}-\frac{5}{2}a\mathcal{J}_{\pm}^{2}\mathcal{J}_{\pm}^{\prime\prime}+\frac{1}{4}b^{2}\mathcal{J}_{\pm}^{6}-\frac{5}{3}b\mathcal{J}_{\pm}^{3}\mathcal{J}_{\pm}^{\prime\prime}+\frac{3}{4}ab\mathcal{J}_{\pm}^{5}+2\mathcal{J}_{\pm}^{\prime\prime2}\right),\nonumber \\
H_{\left(3\right)}^{\pm} & = & \frac{\kappa}{4\pi}\oint d\phi\left(\frac{7}{8}a^{3}\mathcal{J}_{\pm}^{5}-\frac{35}{6}a^{2}\mathcal{J}_{\pm}^{3}\mathcal{J}_{\pm}^{\prime\prime}+7a\mathcal{J}_{\pm}^{2}\mathcal{J}_{\pm}^{\prime\prime\prime\prime}+\frac{5}{16}b^{3}\mathcal{J}_{\pm}^{8}+\frac{35}{2}b^{2}\mathcal{J}_{\pm}^{\prime2}\mathcal{J}_{\pm}^{4}\right.\nonumber \\
&  & \left.+\frac{7}{3}b\left(2\mathcal{J}_{\pm}^{3}\mathcal{J}_{\pm}^{\prime\prime\prime\prime}+\mathcal{J}_{\pm}^{\prime4}\right)+\frac{7}{4}a^{2}b\mathcal{J}_{\pm}^{6}+\frac{5}{4}ab^{2}\mathcal{J}_{\pm}^{7}+35ab\mathcal{J}_{\pm}^{\prime2}\mathcal{J}_{\pm}^{3}-4\mathcal{J}_{\pm}\mathcal{J}_{\pm}^{\prime\prime\prime\prime\prime\prime}\right).\nonumber 
\end{eqnarray}
Note that the Hamiltonians of the Gardner hierarchy cannot be written
as the sum of the Hamiltonians of KdV $(b=0)$ and mKdV $\left(a=0\right)$,
because there are cross terms. The case $H_{\left(-1\right)}^{\pm}$,
that it is obtained by extending the hierarchy backwards, is special
because contains $a^{-1}$, and consequently it cannot be defined
in the mKdV hierarchy.

The generalized Gelfand-Dikii polynomials $R_{\left(k\right)}^{\pm}$
are obtained using eq. \eqref{eq:RHJ}, and take the form

\begin{eqnarray}
R_{\left(-1\right)}^{\pm} & = & \frac{\kappa}{4\pi a},\nonumber \\
R_{\left(0\right)}^{\pm} & = & \frac{\kappa}{4\pi}\mathcal{J_{\pm}},\nonumber \\
R_{\left(1\right)}^{\pm} & = & \frac{\kappa}{4\pi}\left(\frac{3}{2}a\mathcal{J_{\pm}}^{2}+b\mathcal{J_{\pm}}^{3}-2\mathcal{J_{\pm}^{\prime\prime}}\right),\nonumber \\
R_{\left(2\right)}^{\pm} & = & \frac{\kappa}{4\pi}\left(\frac{5}{2}a^{2}\mathcal{J_{\pm}}^{3}-5a\left(\mathcal{J_{\pm}^{\prime}}^{2}+2\mathcal{J_{\pm}}\mathcal{J_{\pm}^{\prime\prime}}\right)+\frac{3}{2}b^{2}\mathcal{J_{\pm}}^{5}\right.\label{eq:k-thR}\\
&  & \left.-10b\left(\mathcal{J_{\pm}}\mathcal{J_{\pm}^{\prime}}^{2}+\mathcal{J_{\pm}}^{2}\mathcal{J_{\pm}^{\prime\prime}}\right)+\frac{15}{4}ab\mathcal{J_{\pm}}^{4}+4\mathcal{J_{\pm}^{\prime\prime\prime\prime}}\right),\nonumber \\
R_{\left(3\right)}^{\pm} & = & \frac{\kappa}{4\pi}\left(\frac{35}{8}a^{3}\mathcal{J}_{\pm}^{4}-35a^{2}\left(\mathcal{J}_{\pm}\mathcal{J}_{\pm}^{\prime2}+\mathcal{J}_{\pm}^{2}\mathcal{J}_{\pm}^{\prime\prime}\right)+7a\left(4\mathcal{J}_{\pm}\mathcal{J}_{\pm}^{\prime\prime\prime\prime}+6\mathcal{J}_{\pm}^{\prime\prime2}+8\mathcal{J}_{\pm}^{\prime}\mathcal{J}_{\pm}^{\prime\prime\prime}\right)\right.\nonumber \\
&  & +\frac{5}{2}b^{3}\mathcal{J}_{\pm}^{7}+7b\left(4\mathcal{J}_{\pm}^{2}\mathcal{J}_{\pm}^{\prime\prime\prime\prime}+12\mathcal{J}_{\pm}\mathcal{J}_{\pm}^{\prime\prime2}+16\mathcal{J}_{\pm}\mathcal{J}_{\pm}^{\prime}\mathcal{J}_{\pm}^{\prime\prime\prime}+20\mathcal{J}_{\pm}^{\prime2}\mathcal{J}_{\pm}^{\prime\prime}\right)-8\mathcal{J}_{\pm}^{\prime\prime\prime\prime\prime\prime}\nonumber \\
&  & \left.-35b^{2}\left(\mathcal{J}_{\pm}^{4}\mathcal{J}_{\pm}^{\prime\prime}+2\mathcal{J}_{\pm}^{3}\mathcal{J}_{\pm}^{\prime2}\right)+\frac{21}{2}a^{2}b\mathcal{J}_{\pm}^{5}+\frac{35}{4}ab^{2}\mathcal{J}_{\pm}^{6}-35ab\left(2\mathcal{J}_{\pm}^{3}\mathcal{J}_{\pm}^{\prime\prime}+3\mathcal{J}_{\pm}^{2}\mathcal{J}_{\pm}^{\prime2}\right)\right).\nonumber 
\end{eqnarray}

\bibliographystyle{JHEP}

\end{document}